\documentclass[aps,prb,onecolumn,floats,superscriptaddress]{revtex4}
\usepackage[pdftex]{graphicx} 
\usepackage{epstopdf}
\usepackage{verbatim}
\usepackage{color}
\usepackage{subfigure}
\usepackage{tabularx}
\usepackage{amsfonts}
\usepackage{wasysym}
\usepackage{bm}
\usepackage[linkcolor=black,anchorcolor=black,citecolor=black,hypertexnames=false]{hyperref}
\usepackage{txfonts}
\usepackage{amssymb}
\usepackage{graphicx}
\usepackage{multirow}
\usepackage{float}

\makeatletter
\def\bib@device#1#2{}
\makeatother

\begin{document}

\large

\title{Structure of spin excitations in heavily electron-doped Li$_{0.8}$Fe$_{0.2}$ODFeSe superconductors}

\author{Bingying Pan$^\sharp$}
\affiliation{State Key Laboratory of Surface Physics and Department of Physics, Fudan University, Shanghai 200433, China}
\author{Yao Shen$^\sharp$}
\affiliation{State Key Laboratory of Surface Physics and Department of Physics, Fudan University, Shanghai 200433, China}
\author{Die Hu}
\affiliation{State Key Laboratory of Surface Physics and Department of Physics, Fudan University, Shanghai 200433, China}
\author{Yu Feng}
\affiliation{State Key Laboratory of Surface Physics and Department of Physics, Fudan University, Shanghai 200433, China}

\author{J. T. Park}
\affiliation{Heinz Maier-Leibnitz Zentrum (MLZ), Technische Universit\"{a}t M\"{u}nchen, D-85748 Garching, Germany}
\author{A. D. Christianson}
\affiliation{Quantum Condensed Matter Division, Oak Ridge National Laboratory, Oak Ridge, Tennessee 37831-6393, USA}
\affiliation{Department of Physics and Astronomy, University of Tennessee, Knoxville, Tennessee 37996, USA}

\author{Qisi Wang}
\affiliation{State Key Laboratory of Surface Physics and Department of Physics, Fudan University, Shanghai 200433, China}
\author{Yiqing Hao}
\affiliation{State Key Laboratory of Surface Physics and Department of Physics, Fudan University, Shanghai 200433, China}
\author{Hongliang Wo}
\affiliation{State Key Laboratory of Surface Physics and Department of Physics, Fudan University, Shanghai 200433, China}

\author{Z. P. Yin}
\affiliation{Department of Physics and Center for Advanced Quantum Studies, Beijing Normal University, Beijing 100875, China}

\author{T. A. Maier}
\affiliation{Computer Science and Mathematics Division and Center for Nanophase Materials Sciences,Oak Ridge National Laboratory, Oak Ridge, Tennessee 37831, USA}
\affiliation{Materials Science and Technology Division,Oak Ridge National Laboratory, Oak Ridge, Tennessee 37831, USA}

\author{Jun Zhao$^\ast$}
\affiliation{State Key Laboratory of Surface Physics and Department of Physics, Fudan University, Shanghai 200433, China}
\affiliation{Collaborative Innovation Center of Advanced Microstructures, Nanjing, 210093, China}



\date{\today}

\begin{abstract}

\end{abstract}


\maketitle

{\bf Heavily electron-doped iron-selenide (HEDIS) high-transition-temperature (high-$T_{\rm{c}}$) superconductors, which have no hole Fermi pockets, but have a notably high $T_{\rm{c}}$, have challenged the prevailing $s$$_\pm$ pairing scenario originally proposed for iron pnictides containing both electron and hole pockets. The microscopic mechanism underlying the enhanced superconductivity in HEDIS remains unclear. Here, we used neutron scattering to study the spin excitations of the HEDIS material Li$_{0.8}$Fe$_{0.2}$ODFeSe ($T_{\rm{c}}$ = 41 K). Our data revealed nearly ring-shaped magnetic resonant excitations surrounding ($\pi$, $\pi$) at $\sim$ 21 meV. As the energy increased, the spin excitations assumed a diamond shape, and they dispersed outward until the energy reached $\sim$ 60 meV and then inward at higher energies. The observed energy-dependent momentum structure and twisted dispersion of spin excitations near ($\pi$, $\pi$) are analogous to those of hole-doped cuprates in several aspects, thus implying that such spin excitations are essential for the remarkably high $T_{\rm{c}}$ in these materials.}

Quantitative knowledge of the energy and momentum dependence of the spin excitations of high-temperature superconductors is essential to establishing the mechanism underlying superconductivity. In iron-pnictide superconductors, it is widely believed that the electron pairing is mediated by the stripe spin fluctuations near the nesting wavevector ($\pi$, 0) between the hole pockets at the Brillouin zone center and the electron pockets at the zone edges (1-Fe unit cell), which typically leads to an $s$-wave pairing with a sign-reversed gap function ($s_{\pm}$) (ref. ~\onlinecite{Dai2015}). However, in the parent phase of iron senlenide superconductors FeSe ($T_{\rm{c}}$ = 8.7 K), in addition to the stripe spin fluctuations near ($\pi$, 0), N\'{e}el spin fluctuations near ($\pi$, $\pi$) were observed over a wide energy range\cite{Wang2016nc}. Moreover, the $s_{\pm}$ pairing scenario is directly challenged by HEDIS because these superconductors contain only electron pockets, but not hole pockets \cite{yanzhang,tqian,dmou,Dagotto2013,He2013,Tan2013,Zhao2016,Niu2015}. In particular, the wavevector connecting the electron pockets is close to ($\pi$, $\pi$) rather than ($\pi$, 0) in HEDIS, and this raises noteworthy questions as to how the spin excitation and pairing symmetry evolve as the system is tuned into the heavily electron-doped regime in which superconductivity is surprisingly enhanced \cite{Wang2012,Ge2015,Guo2010,Dagotto2013}.

The pairing symmetry in HEDIS is currently under intense theoretical debate\cite{Maier2011,fawang,Maiti2011,Das2011,Mazin2011,kontani,pandey,Hao2014,nica,Si2016,Yang2013,Li2016}. In experimental studies, scanning tunneling microscopy quasiparticle interference measurements suggested a sign-preserved $s$-wave superconducting gap function between electron pockets in single-layer FeSe/SrTiO$_3$ thin films (ref.~\onlinecite{Fan2015}), whereas, low-energy inelastic neutron scattering data revealed a magnetic resonant mode around ($\pi$, 0.5$\pi$) in $A_x$Fe$_{2-y}$Se$_2$ ($A$ = K, Rb...), indicating a sign-reversed superconducting gap function \cite{Park2011,Taylor2012,Friemel2012}. However, the lack of phase-pure single-crystalline samples has rendered impractical the measurement of the momentum structure of spin excitations over a wider energy range throughout the entire Brillouin zone in any HEDIS superconductor. In addition, studies have argued that the coexisting $\sqrt{5}\times\sqrt{5}$ iron vacancy ordered antiferromagnetic insulating phase in $A_x$Fe$_{2-y}$Se$_2$ and the substrate of the single-layer FeSe/SrTO$_3$ may influence superconductivity \cite{Li2012,Lee2014}. As yet a complete picture of the spin excitations and the pairing symmetry of a pure bulk single-crystalline HEDIS material remain unclear.

The newly discovered HEDIS Li$_{0.8}$Fe$_{0.2}$OHFeSe ($T_c=41$ K) exhibits remarkably similar electronic and superconducting gap structures to those of single layer FeSe/SrTO$_3$ (refs \onlinecite{Lu2014,Pachmayr,Niu2015,Zhao2016}). Particularly, because the phase-pure single-crystalline Li$_{0.8}$Fe$_{0.2}$OHFeSe with a sufficiently large size can be grown, the corresponding intrinsic bulk properties not subject to the influence of the interface or impurity phases can be investigated.

In this paper, we report neutron scattering measurements of the spin excitations over a wide range of momentum and energy in single-crystalline Li$_{0.8}$Fe$_{0.2}$ODFeSe ($T_{\rm{c}}$ = 41 K) (Methods). Our data revealed nearly ring-shaped magnetic resonant excitations at $\sim$ 21 meV, comprising four elliptical peaks at ($\pi$, $0.62\pi$) and equivalent wavevectors surrounding ($\pi$, $\pi$). The presence of the resonance mode surrounding ($\pi$, $\pi$) indicates a sign reversal in the superconducting gap function between the electron pockets. As the energy increased, the spin excitations assumed a diamond shape, and they dispersed outward until the energy reached $\sim$ 60 meV and then inward at higher energies, eventually forming a big blob near ($\pi$, $\pi$) at 130 meV. Such an energy-dependent momentum structure and twisted dispersion of spin excitations near ($\pi$, $\pi$) resemble to those in hole-doped cuprates in several aspects\cite{Hayden,Tranquada2004,Vignolle2007,Lipscombe}. Our results imply that such spin excitations are an essential ingredient for high temperature superconductivity in these materials.

{\bf Results}

{\bf Magnetic resonance mode.}

 We first used the PUMA thermal triple-axis spectrometer to measure low energy spin excitations and their interplay with superconductivity. Figure 1a illustrates the energy scans at $\bf{Q}$=(0.5, 0.69, 0) in the superconducting state ($T$ = 2.6 K) and normal state ($T$ = 45 K), indicating that the scattering at $\sim$ 21 meV is enhanced below $T_{\rm{c}}$, whereas that at lower energies is suppressed below $T_{\rm{c}}$. To ensure a clear illustration of the effect of superconductivity, Fig. 1b presents the plot of the intensity difference associated with the energy scans between the superconducting state and normal state, revealing a spin-gap like feature below 18 meV and a resonance mode at $\sim$ 21 meV below the superconducting gap (2$\Delta$ $\approx$ 28 meV) (refs \onlinecite{Zhao2016,Du2016}). A similar resonance mode at $\sim$ 20 meV is also revealed near $\bf{Q}$=(0.5, 0.31, $L$) in the data measured on the ARCS time-of-flight (TOF) chopper spectrometer (Fig. 1c). This is consistent with a spin exciton within the superconducting gap when the gap function exhibited an opposite sign between the electron pockets at the adjacent zone edges \cite{Maier2011}.

The resonance excitation at 21 meV is further confirmed by the {\bf{Q}}-scans near (0.5, 0.69, 0) across $T_{\rm{c}}$. Figure 1d shows that the normal-state spin excitation ($T = 45$ K) can be described by a single Gaussian peak on a linear background and that it is significantly enhanced on entering the superconducting state ($T = 2.6$ K). Figure 1g illustrates the detailed temperature dependence of the scattering, again confirming that the enhancement of the spin excitations is tightly associated with superconductivity and behaves like an order parameter below $T_{\rm{c}}$.

To precisely determine the momentum structure of the resonance mode, we used the ARCS spectrometer to map out the Brillouin zone (Methods). The intensity difference image between the superconducting state and normal state [$S$(5 K)-$S$(50 K)] near the resonance energy (Fig. 1h) showed elliptical peaks at ($0.5$, $0.5\pm\delta$) and ($0.5\pm\delta$, $0.5$) symmetrically surrounding (0.5, 0.5), forming a nearly ring-like structure. The peak position (0.5, 0.5$\pm$0.19) and the anisotropy of the scattering in momentum space were further confirmed by constant-energy cuts, revealing a significant difference between the peak widths along the $H$ and $K$ directions (Fig. 1e,f). We noted that the observed incommensurability $\delta$=0.19 was smaller than that observed ($\delta$ $\approx0.25$) for the resonance mode in $A_x$Fe$_{2-y}$Se$_2$ (refs ~\onlinecite{Park2011,Taylor2012,Friemel2012}). This is probably because the electron doping level in Li$_{0.8}$Fe$_{0.2}$ODFeSe ($\sim$0.08-0.1 electrons per Fe) is lower (less overdoped) than that in $A_x$Fe$_{2-y}$Se$_2$ ($\sim$0.18 electrons per Fe) (ref. ~\onlinecite{Zhao2016}). Therefore, the low-energy magnetic scattering in Li$_{0.8}$Fe$_{0.2}$ODFeSe is closer to ($\pi$, $\pi$) compared with that in $A_x$Fe$_{2-y}$Se$_2$ (Fig. 1h, 1i).

{\bf Momentum and energy dependence of high energy spin excitations.}

To acquire a complete picture of the spin excitations, we present in Fig. 2 the evolution of the spin response with energy. At low energies, the momentum structure of the spin excitations was similar to that of the resonance mode (Fig. 2a); as the energy increased, the magnetic response assumed a diamond shape and dispersed outward (Fig. 2b-d). Concurrently, the major axis of the elliptical peaks was rotated by 90$^\circ$ at energies of $\sim$ 59-66 meV with respect to the axis of those observed at lower energy (marked by dashed ellipses in Fig. 2a and 2d; the rotation of the elliptical peaks is further illustrated in Supplementary Fig. 1). When the energy exceeded 66 meV, the scattering dispersed inward and formed a nearly ring-like pattern at 100 meV, and eventually becoming a big blob near (0.5, 0.5) at 130 meV (Fig. 2e-j).  We note that a weak ferromagnetic peak near Q $\approx$ 0, which likely arises from the Li-Fe layer, was revealed by small angle neutron scattering measurements in polycrystalline ($^7$Li$_{0.82}$Fe$_{0.18}$OD)FeSe ($T_{\rm{c}}$ = 18 K) (ref. ~\onlinecite{Lynn2015}). This wavevector (Q $\approx$ 0) was not covered in our measurements in the first Brillouin zone. We also did not observe clear ferromagnetic excitations in the second Brillouin zone, which could be due to the intrinsic weak signal and the decreased magnetic form factor. It is also possible that our sample which has a different chemical composition and a higher $T_{\rm{c}}$ (41 K) has weaker or no ferromagnetic correlations.

The overall dispersion of the spin excitations in Li$_{0.8}$Fe$_{0.2}$ODFeSe can be seen more clearly in $E$-$\bf{Q}$ space. As the energy increased, the magnetic excitations dispersed outward and then inward (Fig. 3a,b). This dispersion is further confirmed by the constant-energy cuts along the $K$ direction at various energies in Fig. 3c. Such a strongly energy-dependent momentum structure and twisted dispersions of spin excitations are in analogy to those of hole-doped cuprate superconductors, which typically demonstrate the existence of inflection/saddle points in the dispersion curves and the rotation of the scattering pattern across inflection/saddle points\cite{Tranquada2004,Hayden,Vignolle2007}. On the other hand, spin excitations in most iron-pnictide superconductors generally disperse outward from ($\pi$, 0) to ($\pi$, $\pm\pi$) which can usually be described by either an itinerant or local moment model \cite{Dai2015}, or the density functional theory (DFT) combined with dynamical mean field theory (DMFT) taking into account both correlation effects and realistic band structures \cite{Yin2014}. It should be noted that the magnetically-ordered nonsuperconducting Fe$_{1+x}$Te also displays unusual hourglass-like dispersions \cite{Stock2014,Igor}. However, the magnetic interaction in Fe$_{1+x}$Te is very complicated because of the spiral magnetism induced by the interstitial Fe (ref. ~\onlinecite{Stock2014}) and the competition between the double stripe ($\pi/2$, $\pi/2$) and stripe ($\pi$, 0) magnetism \cite{Yin2014,Yin2012}.

{\bf Momentum integrated local susceptibility}

The use of phase-pure single crystals enables executing quantitative spin excitation measurements in absolute units, which has not been feasible for phase-separated $A_x$Fe$_{2-y}$Se$_2$. Figure 4 shows the energy dependence of the local dynamic susceptibility of Li$_{0.8}$Fe$_{0.2}$ODFeSe in absolute units, revealing a two-peak structure, which is similar to those of iron pnictides and FeSe \cite{Dai2015,LiuM2012,Wang2013,Wang2016nc}, although the momentum structures of the spin excitations in these materials are quite different. The peak of the lower energy component corresponds to the resonance mode; the higher energy component, which carries much more spectral weight, peaked near 60-110 meV (Fig. 4). Notably, the integrated resonance spectral weight [$\chi^{''}_{5K}-\chi^{''}_{50K}$ = 0.029(7) $\mu_{\rm{B}}^2$ Fe$^{-1}$] in Li$_{0.8}$Fe$_{0.2}$ODFeSe is at least one order of magnitude larger than that in bulk FeSe ($T_{\rm{c}}$ = 8.7 K). However, the high energy component of the spectral weight is much lower than that in FeSe (ref.~\onlinecite{Wang2015,Wang2016nc}). The effect of the electron-doping seems to entail the considerable suppression of high-energy responses and enhancement of low-energy responses. This behavior resembles that of hole-doped cuprates and hole-doped iron pnictides\cite{Vignolle2007,Wang2013}, but differs from that of the electron-doped iron pnictides in which the high energy spin excitations were essentially doping independent and the superconductivity was completely suppressed in the over electron-doped regime\cite{Wang2013}. The spin excitation band widths of Li$_{0.8}$Fe$_{0.2}$ODFeSe and FeSe are lower than those of most iron pnictides, indicating stronger electron correlations in FeSe-based superconductors \cite{Yin2014}.

{\bf BCS/RPA calculations and DFT+DMFT calculations}

The magnetism in iron-based materials may arise from either the exchange interaction of the local moments or Fermi surface nesting of itinerant electrons, or both\cite{Dai2015}. To the best of our knowledge, the observed energy-dependent momentum structure and twisted dispersion of the spin excitations have not been predicted in the known theoretical calculations in HEDIS based on either a local or itinerant picture. Using a BCS/RPA approach \cite{Maier2011}, we have calculated the magnetic susceptibility from a 2D tight-binding five-orbital Hubbard-Hund Hamiltonian that describes the electronic structure of an FeSe system with only electron pockets \cite{Maier2011}. For the superconducting gap we have used a phenomenological $d_{x^2-y^2}$ gap $\Delta(k) = \Delta_0(\cos k_x - \cos k_y)$, which is close to isotropic on the electron pockets and changes sign between them. Previous calculations have shown that this type of gap structure leads to a neutron resonance in $\chi¡¯¡¯(q,\omega)$ near $q\approx(\pi,0.6\pi)$ that arises from scattering between the electron pockets on which the gap changes sign \cite{Maier2011,friemel}. The interaction matrix in orbital space used in the RPA calculation contains on-site matrix elements for the intra- and inter-orbital Coulomb repulsions $U$ and $U^\prime$ and for the Hunds-rule coupling and pair-hopping terms $J$ and $J^\prime$, respectively. Here we have used spin-rotationally invariant parameters $J=J^\prime=U/4$ and $U^\prime=U/2$ with $U=0.96 {\rm eV}$. Fig. 5a shows the calculated dispersion of spin excitations as the energy increases and Fig. 5b-e displays its momentum structure for different energies. As one sees, the momentum position of the magnetic excitations below 40 meV is broadly consistent with the experiments (Fig. 3a, 3b and 5a). However, the BCS/RPA calculations fail to describe the outward dispersion at higher energies seen in Fig. 3a,b as well as the twisted momentum structure seen in Fig. 2. We also used a combination of density functional theory and dynamical mean field theory, so called DFT+DMFT as implemented in ref.~\onlinecite{Yin2014} to compute the electronic structure and spin dynamics of this compound (Supplementary Fig. 2; Supplementary Note 1). Similar to the BCS/RPA calculations, the DFT+DMFT-calculated spin excitation spectra also only show inward dispersion, and no twisted structure is observed (Supplementary Fig. 2i).

{\bf Discussion}

The distinct dispersion of spin excitations below and above $\sim 60$ meV might imply different origins of these excitations. The low-energy spin excitations are possibly related to Fermi surface nesting, as the resonance wavevector and its doping dependence agree with our BCS/RPA calculations \cite{Maier2011,Zhao2016,Park2011}; whereas, the high-energy spin excitations, which carry more spectral weight, are possibly due to vestigial short-range magnetic exchange interactions, as also observed in hole-doped cuprates \cite{Vignolle2007,Lipscombe,Kivelson2003}. Regardless of their origins, the spin excitations surrounding ($\pi$, $\pi$) as well as the electron Fermi pockets connected by ($\pi$, $\pi$) may interact collaboratively to enhance superconductivity. These structures of spin excitations and Fermi surfaces are analogous to those of hole-doped cuprates, thus implying that such spin excitations are a key ingredient for the remarkably high $T_{\rm{c}}$ in these materials.

\textit{Note added:} Recently, we noticed a related preprint describing low-energy spin excitations on Li$_{1-x}$Fe$_{x}$ODFe$_{1-y}$Se powder samples \cite{Davies2016}.

\textbf{Methods}

\textbf{Sample growth and characterizations.} Li$_{0.8}$Fe$_{0.2}$ODFeSe single crystals were grown using a hydrothermal method similar to that described in ref.~\onlinecite{Dong2015} and were deuterated to reduce the incoherent scattering from hydrogen atoms for the inelastic neutron scattering measurements. Magnetic susceptibility and resistivity measurements conducted on a crystal from the same batch as those measured with neutrons showed a sharp superconducting transition at 41 K, signifying the high quality of the crystal. The single crystalline sample was also ground into powder for the X-ray diffraction refinements (Supplementary Fig. 3; Supplementary Note 2; Supplementary Table 1). The refined structure parameters are consistent with those in previous reports\cite{Lu2014}.

\textbf{Inelastic neutron scattering experiments.} Inelastic neutron scattering measurements were carried out on the PUMA thermal neutron triple-axis spectrometer (TAS) at the Heinz Maier-Leibnitz Zentrum (MLZ), TU M$\rm{\ddot{u}}$nchen, Germany, and the ARCS time-of-flight (TOF) chopper spectrometer at the Spallation Neutron Source of Oak Ridge National Laboratory, USA (Supplementary Note 3). For the TAS experiment, approximately 30 pieces of single crystals with a total mass of 3.2 g were coaligned in the ($H$ $K$ 0) plane to a mosaicity within $\sim4^\circ$. For the TOF experiment, 8.5 g crystals were coaligned to a masaicity within $\sim5^\circ$; the incident beam was parallel to the $c$-axis. The wavevector $\bf{Q}$ at ($q_x$, $q_y$, $q_z$) is defined as ($H$, $K$, $L$) = ($q_xa/2\pi$, $q_ya/2\pi$, $q_zc/2\pi$) reciprocal lattice units (r.l.u.) in the 1-Fe unit cell. Here, (0.5, 0.5) and (0.5, 0) correspond to ($\pi$, $\pi$) and ($\pi$, 0), respectively. The $\vert$$\bf{Q}$$\vert$-dependent background was subtracted for the data in Fig. 2a-e following the practice of ref. 2 (Supplementary Fig. 4; Supplementary Note 4). Additional low energy data are shown in Supplementary Fig. 5 (Supplementary Note 5). The scattering intensity was normalized into absolute units by calibrating it against the incoherent scattering of a standard vanadium sample.

{\bf Data availability.} All data that support the findings of this study are available from the corresponding authors upon request.

$^{\sharp}$These authors contributed equally to this work.

$^{*}$Correspondence and requests for materials should be addressed to J.Z. (zhaoj@fudan.edu.cn).

{\bf Acknowledgements}

This work was supported by the National Natural Science Foundation of China (Grant No. 11374059), the National Key R\&D Program of the MOST of China (Grant No. 2016YFA0300203) and the Ministry of Science and Technology of China (Program 973: 2015CB921302). Z.P.Y. was supported by the National Natural Science Foundation of China (Grant No. 11674030) and the National Key Research and Development Program of China (contract No. 2016YFA0302300). T.A.M. was supported by the U.S. Department of Energy, Office of Basic Energy Sciences, Materials Sciences and Engineering Division. A portion of this research used resources at the Spallation Neutron Source, a DOE Office of Science User Facility operated by the Oak Ridge National Laboratory.

{\bf Author contributions}

 J.Z. planned the research. D.H, Y.F., and B.P. synthesized the sample. B.P., D.H., Y.S., Q.W., Y.H., and H.W. characterized the sample. B.P. and Y.S. conducted the neutron experiments with the experimental assistance from J.T.P. and A.D.C.. T.A.M. did the BCS/RPA calculations. Z.P.Y. performed DFT+DMFT calculations. J.Z., B.P. and Y.S. analysed the data and wrote the paper.

\textbf{Additional information}

The authors declare no competing financial interests. Supplementary information
accompanies this paper at www.nature.com. Reprints and permission information is available online
at http://www.nature.com/reprints.

\textbf{References}

\newpage

\begin{figure*}[b]

\includegraphics[clip,width=16cm]{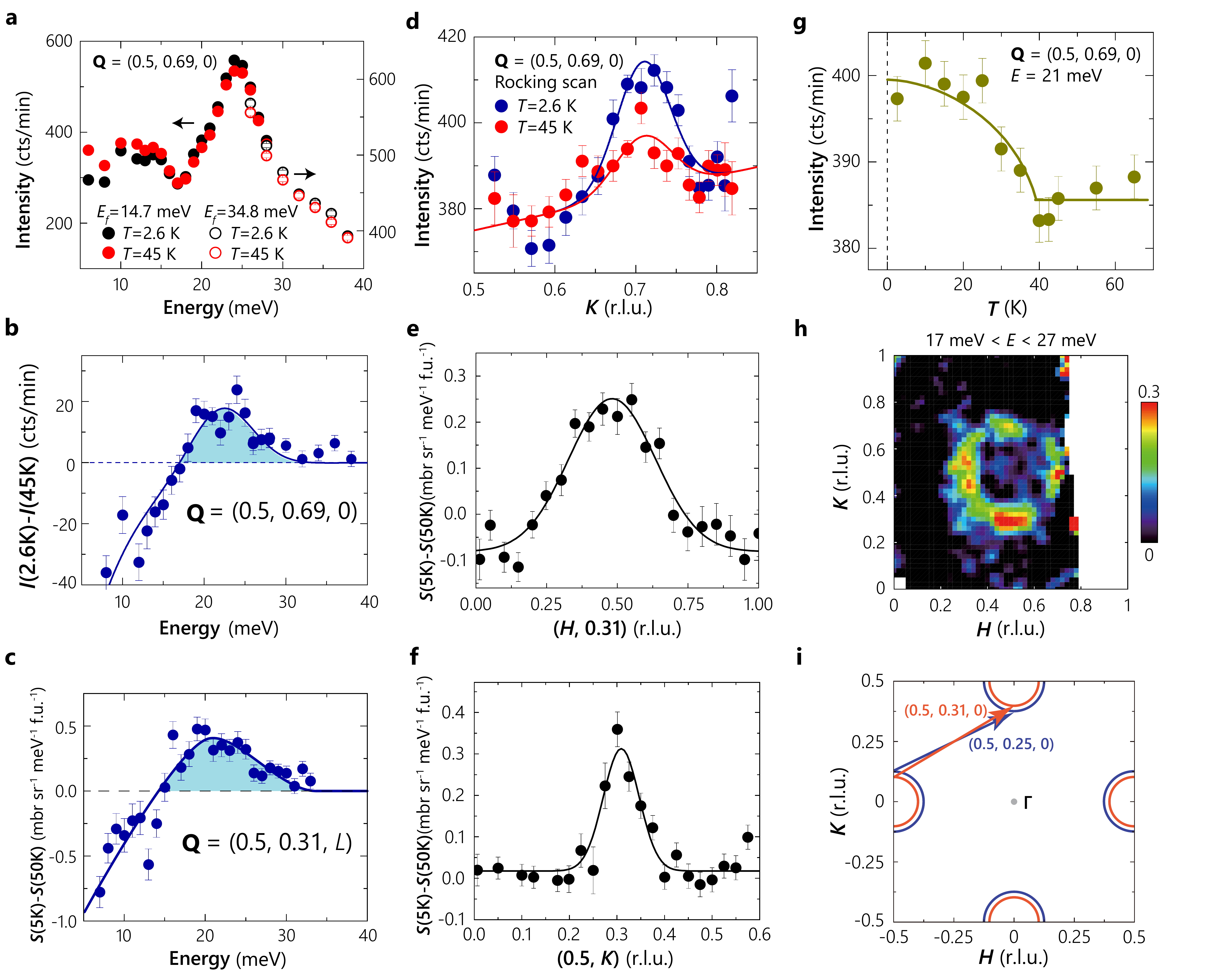}
\caption{\label{fig:epsart} \textbf{Magnetic resonant mode in Li$_{0.8}$Fe$_{0.2}$ODFeSe ($T_{\rm{c}}$ = 41 K).} $\mathrm{\bf{a}}$, Energy dependence of spin excitations for Li$_{0.8}$Fe$_{0.2}$ODFeSe at $\bf{Q}$ = (0.50, 0.69, 0) in the superconducting state (T = 2.6 K) and normal state (45 K). The solid and open circles correspond to the data collected at final energies of $E_f$ = 14.7 and 34.8 meV, respectively. $\bf{b}$, Intensity difference between the superconducting state and normal state [$S$(2.6K)-$S$(45K)] at (0.50, 0.69, 0) measured on the PUMA thermal triple-axis spectrometer. The data collected at different $E_f$ were normalized. $\bf{c}$, Intensity difference between the superconducting state and normal state [$S$(5 K)-$S$(50 K)] at (0.50, 0.31, $L$) measured on the ARCS time of flight spectrometer ($0.5\le L\le 4$). $\bf{d}$, Rocking scan near (0.50, 0.69, 0) at $E$ = 21 meV at T=2.6 K and 45 K. $\bf{e}$,$\bf{f}$, Intensity difference between the superconducting state and normal state [$S$(5K)-$S$(50K)] along the ($H$, 0.31) and (0.50, $K$) directions. $\bf{g}$, Temperature dependence of the scattering at (0.50, 0.69, 0) and $E$ = 21 meV. $\bf{h}$, Intensity difference image [$S$(5K)-$S$(50K)] at $E$ = 22 $\pm$ 5 meV. The color bar indicates this intensity difference in unit of mbr sr$^{-1}$ meV$^{-1}$ f. u.$^{-1}$. The white regions in the color plot are gaps between neutron detectors. $\bf{i}$, Schematic of the electron Fermi pockets in Li$_{0.8}$Fe$_{0.2}$ODFeSe (red) and (Tl,Rb)$_x$Fe$_{2-y}$Se$_2$ (blue) adapted from ref.~\onlinecite{Zhao2016}. The calculated resonance wavevector is approximately (0.50, 0.3125) for 0.1 electrons per Fe (ref.~\onlinecite{Maier2011}), which is consistent with our data (0.50, 0.31). The error bars indicate one standard deviation.}

\end{figure*}

\begin{figure*}[t]
\includegraphics[clip,width=18cm]{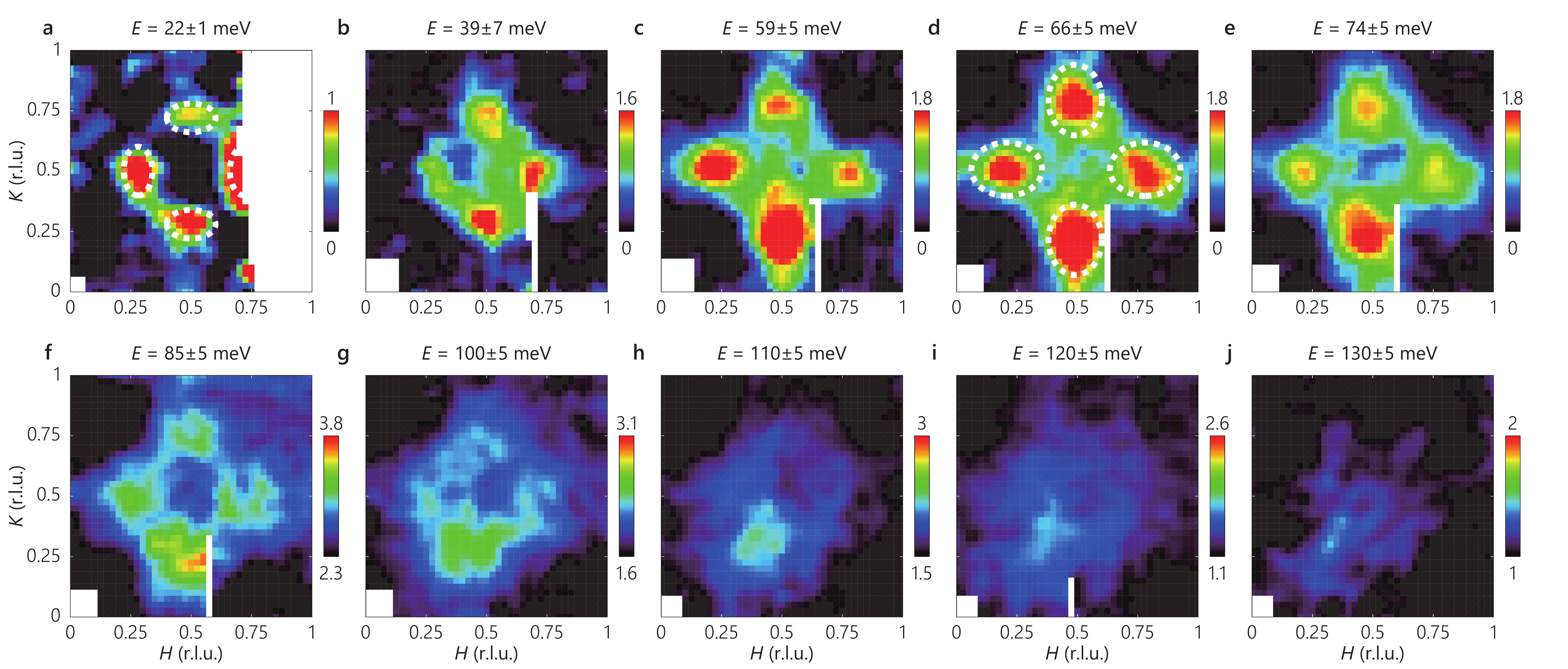}
\caption{\label{fig:epsart} \textbf{Momentum dependence of the spin excitations in Li$_{0.8}$Fe$_{0.2}$ODFeSe at $T$ = 5 K}. $\bf{a-j}$, Constant-energy images acquired at 5 K and at indicated energies. $|\bf{Q}|$-dependent background was subtracted in ($\bf{a}$-$\bf{e}$) (see Supplementary Note 4). For energies $\ge$ 85 meV, raw data are presented ($\bf{f}$-$\bf{j}$).  The measurements in ($\bf{a}$) and ($\bf{b}$-$\bf{j}$) were conducted at the incident neutron energies of 49.6 and 191.6 meV, respectively. Symmetry equivalent data were collected and averaged to enhance statistical accuracy. The color bar indicates scattering intensity in unit of mbr sr$^{-1}$ meV$^{-1}$ f. u.$^{-1}$.}
\end{figure*}

\begin{figure*}[t]
\includegraphics[clip,width=16cm]{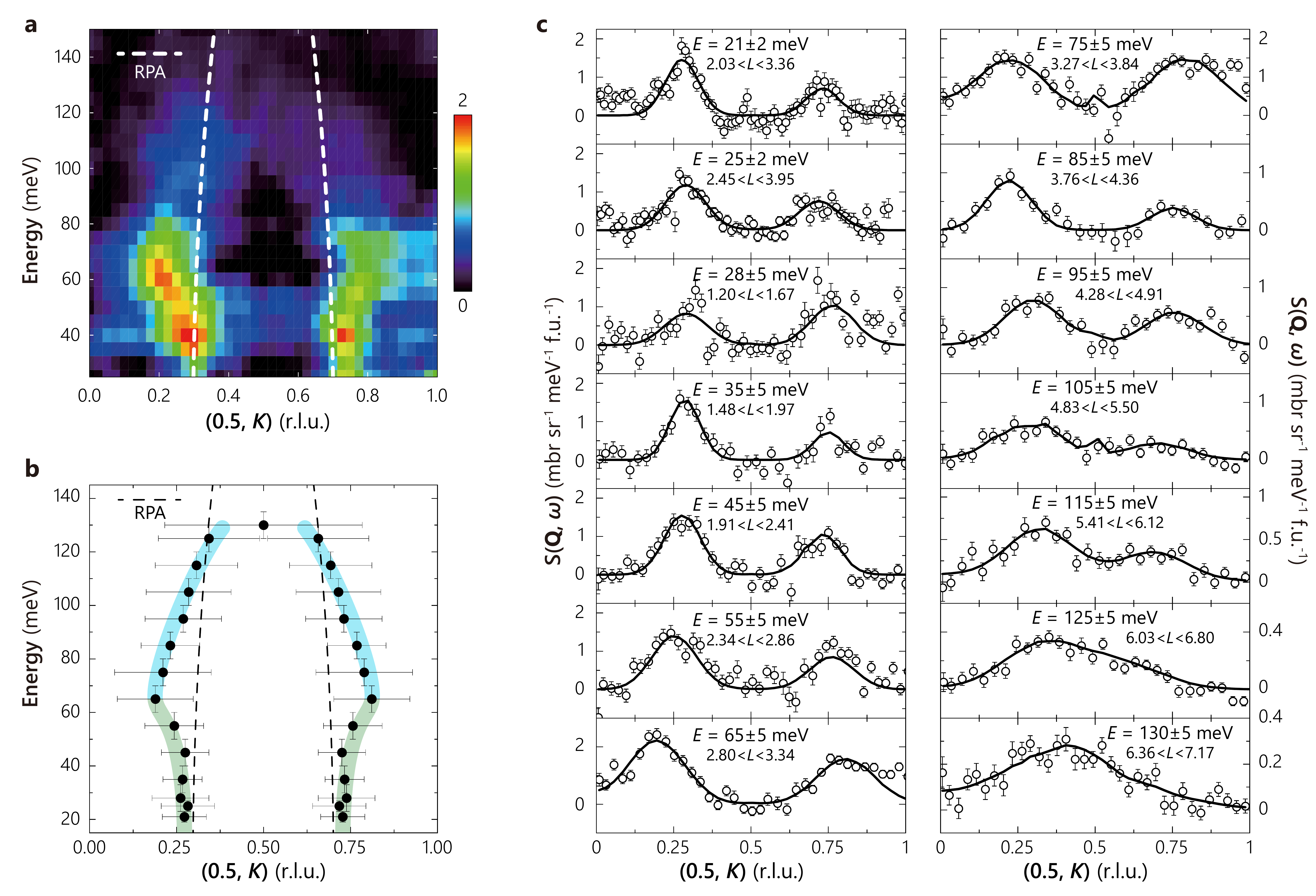}
\caption{\label{fig:epsart} \textbf{Dispersion of the spin excitations in Li$_{0.8}$Fe$_{0.2}$ODFeSe at $T$ = 5 K.} $\bf{a}$, Background subtracted $E$-$K$ slice with incident neutron energy of 191.6 meV. A twisted dispersion is clearly illustrated. The incident neutron beam was parallel to the $c$ axis and $L$ was coupled with the energy transfer. No $L$ modulations were observed from the scattering, indicating a two-dimensional nature of the magnetism. The color bar indicates scattering intensity in unit of mbr sr$^{-1}$ meV$^{-1}$ f. u.$^{-1}$. $\bf{b}$, Dispersion acquired from the constant energy cuts in ($\bf{c}$). The horizontal error bars represent the full-width at half-maximum of the Gaussian peaks in ($\bf{c}$); the vertical error bars represent the energy integration interval. The curves are guides to the eye. $\bf{c}$, Constant energy cuts along the (0.5, $K$) direction at the indicated energies and energy/$L$ integration interval. The measurements were conducted at incident neutron energies of 49.6 and 191.6 meV. The peak positions are determined by fitting with Gaussian profiles convoluted with the instrumental resolution, with a correction of the Fe$^{2+}$ magnetic form factor. The error bars indicate one standard deviation. We note that the magnetic intensities in ($\bf{a}$, $\bf{c}$) and Fig. 2 are asymmetric because the magnetic form factor drops at larger $|\bf{Q}|$.}
\end{figure*}

\begin{figure*}[]
\includegraphics[clip,width=14cm]{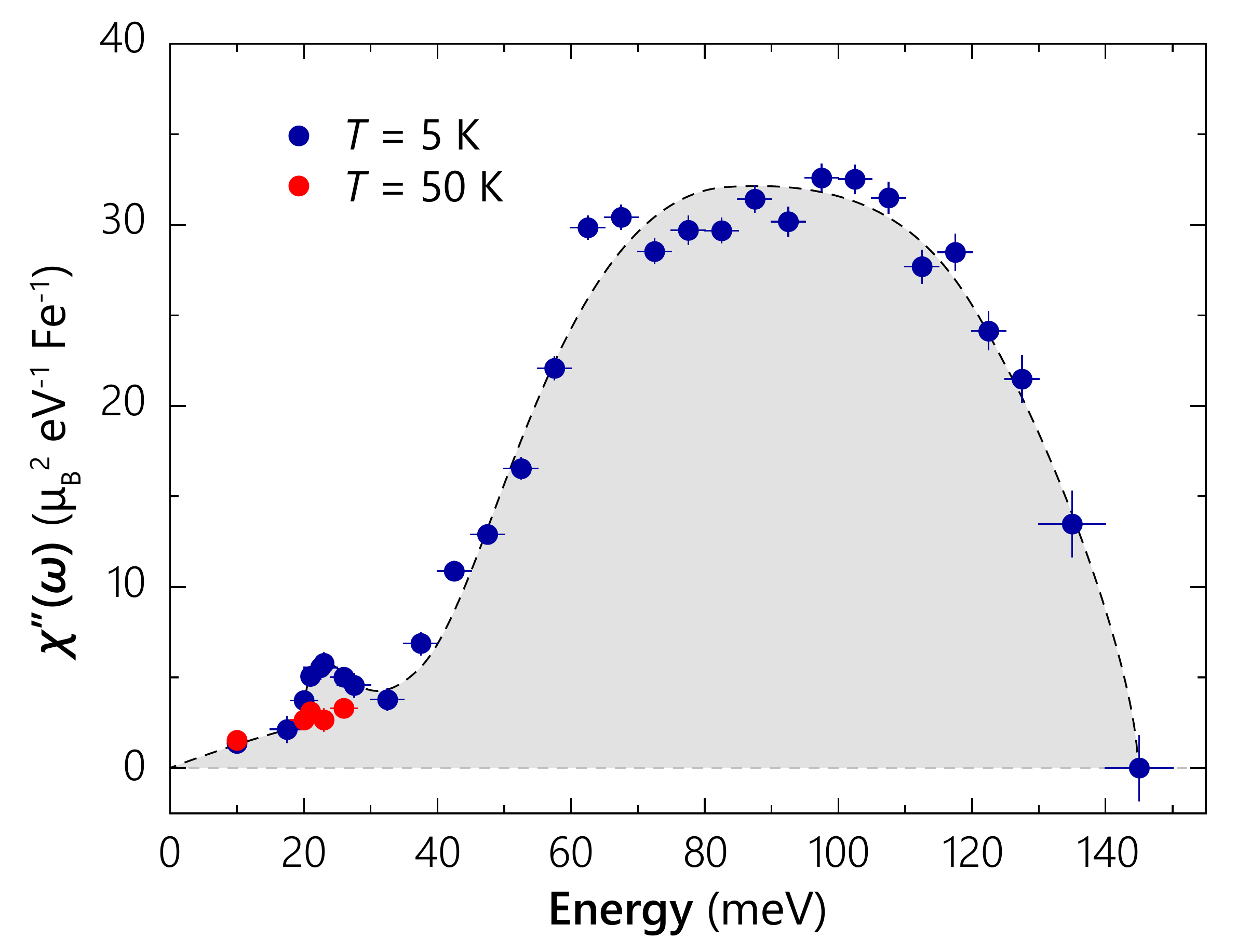}
\caption{\label{fig:epsart} \textbf{Energy dependence of the dynamic local susceptibility of Li$_{0.8}$Fe$_{0.2}$ODFeSe at T = 5 and 50 K.} The vertical and horizontal bars represent the standard deviation and integrated energy interval, respectively. The dashed curves are guides to the eye.}
\end{figure*}

\begin{figure*}[]
\includegraphics[clip,width=10cm]{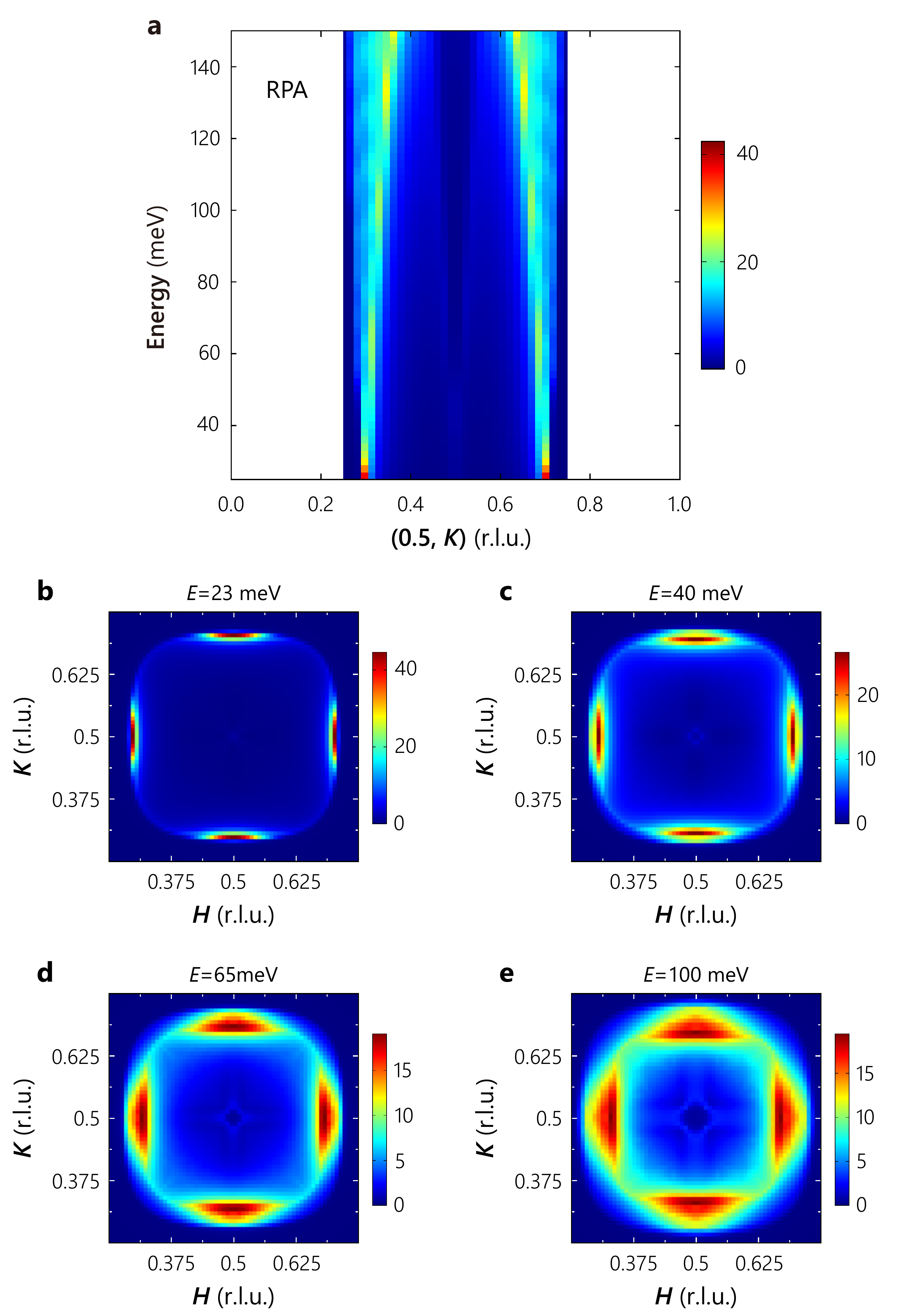}
\caption{\label{fig:epsart} \textbf{Calculated dispersion of the spin excitations of Li$_{0.8}$Fe$_{0.2}$ODFeSe in the superconducting state.} The
calculation was done using a BCS/RPA approach described in the main text. $\bf{a}$, Dispersion of the spin excitations. $\bf{b-e}$, Momentum structure of the spin excitations at indicated energies. The color bars indicate intensity in arbitrary unit.}
\end{figure*}

\end{document}